\documentclass[a4paper]{article}

\usepackage{INTERSPEECH2021}
\usepackage{graphicx}
\usepackage{float}
\usepackage{booktabs}
\usepackage{xcolor}
\usepackage{color}
\usepackage{hyperref}

\usepackage{subcaption}

\usepackage{caption}
\usepackage{enumitem}

\title{{DelightfulTTS 2: End-to-End Speech Synthesis with Adversarial Vector-Quantized Auto-Encoders}}
\name{Yanqing Liu$^{1*}$, Ruiqing Xue$^{1*}$, Lei He$^1$, Xu Tan$^2$, Sheng Zhao$^1$}
\address{
  $^1$Microsoft Azure Speech\\
  $^2$Microsoft Research Asia}
\email{\{yanqliu, ruiqingxue, helei, xuta, szhao\}@microsoft.com}
\begin{document}

\maketitle
\def\thefootnote{*}\footnotetext[1]{These authors contributed equally to this work. Corresponding author: Yanqing Liu, yanqliu@microsoft.com}\def\thefootnote{\arabic{footnote}}
%\def\thefootnote{*}\footnotetext[1]{Corresponding author: Yanqing Liu, yanqliu@microsoft.com}\def\thefootnote{\arabic{footnote}}
%\footnote{normal footnote}

\begin{abstract}
Current text to speech (TTS) systems usually leverage a cascaded acoustic model and vocoder pipeline with mel-spectrograms as the intermediate representations, which suffer from two limitations: 1) the acoustic model and vocoder are separately trained instead of jointly optimized, which incurs cascaded errors; 2) the intermediate speech representations (e.g., mel-spectrogram) are pre-designed and lose phase information, which are sub-optimal. To solve these problems, in this paper, we develop DelightfulTTS 2, a new end-to-end speech synthesis system with automatically learned speech representations and jointly optimized acoustic model and vocoder. Specifically, 1) we propose a new codec network based on vector-quantized auto-encoders with adversarial training (VQ-GAN) to extract intermediate frame-level speech representations (instead of traditional representations like mel-spectrograms) and reconstruct speech waveform; 2) we jointly optimize the acoustic model (based on DelightfulTTS) and the vocoder (the decoder of VQ-GAN), with an auxiliary loss on the acoustic model to predict intermediate speech representations. Experiments show that DelightfulTTS 2 achieves a CMOS gain +0.14 over DelightfulTTS, and more method analyses further verify the effectiveness of the developed system.

\end{abstract}
%\end{}

\section{Introduction}
Popular TTS models~\cite{tan2021survey} are based on a typical two-stage system consisting of an acoustic model and a vocoder. The phonemes or linguistic features from input text are transformed into intermediate acoustic representations like mel-spectrograms by an acoustic model and the predicted acoustic representations are then converted to waveform with a vocoder. Although two-stage TTS systems~\cite{li2019neural,ren2021fastspeech,liu2021delightfultts} have been showing good quality in terms of prosody and audio fidelity in the past years, they still suffer several issues: 1) Mel-spectrograms are extracted by Fourier transformation where phase information are lost, which are not an optimal representation to cascade acoustic models and vocoders together~\cite{cong2021glow}. 2) Since vocoders are trained with ground-truth mel-spectrograms while mel-spectrograms predicted by acoustic models are used in inference, the inaccurate predictions from acoustic models would cause training-inference mismatch and thus inferior audio quality.

To solve the issues in two-stage cascaded TTS systems, fully end-to-end TTS models~\cite{donahue2020end, ren2021fastspeech,kim2021conditional,weiss2021wave,tan2022naturalspeech} are developed recently. Although ideally having advantages over two-stage systems, different end-to-end models face their own challenges and limitations: 1) no significant improvement over two-stage models in terms of voice quality, such as FastSpeech 2s~\cite{ren2021fastspeech}, EATS~\cite{donahue2020end}, and WAVE-TACOTRON~\cite{weiss2021wave}; 2) slow inference speed due to autoregressive generation, such as WAVE-TACOTRON; 3) complicated training pipeline, such as VITS~\cite{kim2021conditional} and ClariNet~\cite{ping2018clarinet}; 4) relying on Fourier transform for representation extraction, such as mel-spectrogram FastSpeech 2s~\cite{ren2021fastspeech} and linear spectrogram in VITS~\cite{kim2021conditional}.  While mel-spectrogram or linear spectrogram is already a compact representation of speech, it still lacks some speech details \cite{cong2021glow}.

In this paper, we develop DelightfulTTS 2, a new end-to-end speech synthesis system with automatically learned frame-level speech representations and jointly optimized acoustic model and vocoder. Specifically, we introduce the designs in DelightfulTTS 2 as follows. 
\begin{itemize}[leftmargin=*]
\item We propose a new codec network based on vector-quantized auto-encoders with adversarial training (VQ-GAN) to automatically learn intermediate frame-level speech representations, instead of using mel-spectrograms or other pre-designed features. Specifically, we use the encoder in VQ-GAN to extract speech representations, and quantize them with multi-stage vector quantizers, and then use the decoder to reconstruct waveform with adversarial training.
\item After the VQ-GAN is trained, we jointly optimize the vocoder (i.e., the decoder of VQ-GAN) with an acoustic model based on DelightfulTTS~\cite{liu2021delightfultts}, with an auxiliary loss on the acoustic model to predict the learned intermediate speech representations extracted by the encoder of VQ-GAN. 
\end{itemize}

We conduct experiments on an internal English dataset. Experiment results show that DelightfulTTS 2 achieves +0.14 CMOS gain over the baseline DelightfulTTS and further objective evaluations also verify its effectiveness for TTS modelling.

\begin{figure*}[h]
  \centering
  \includegraphics[width=0.8\textwidth,trim=0.5cm 0.5cm 0.5cm 0.5cm,clip=true]{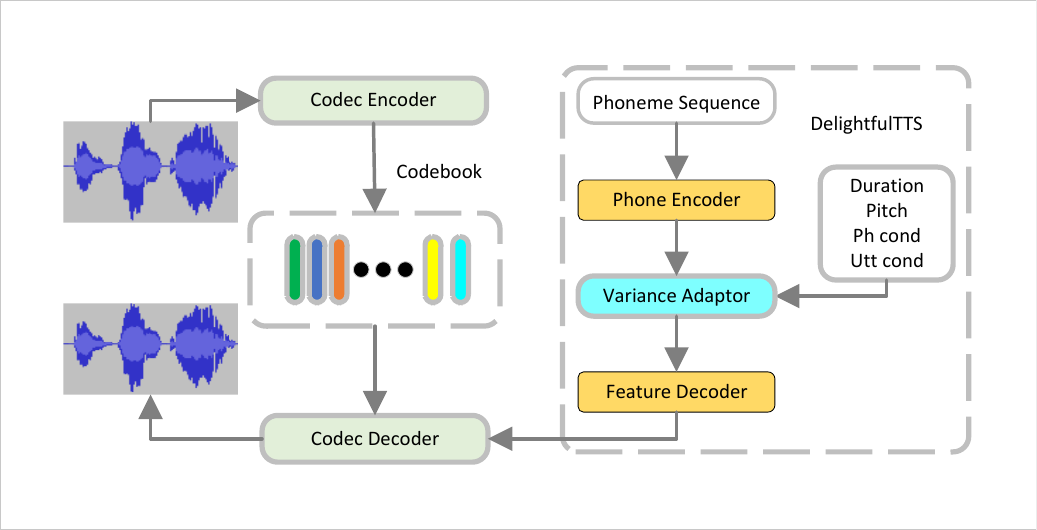}
  \caption{Overview of our proposed DelightfulTTS 2. The left is a codec network for speech representation learning and waveform reconstruction, whose encoder acts like a feature extractor and decoder acts like a vocoder. The right is an acoustic model based on DelightfulTTS that maps phoneme sequence into learnt speech representations, ``Ph cond'' is phoneme-level acoustic condition while ``Utt cond'' is utterance-level acoustic condition. The vocoder and acoustic model are jointly optimized.}
  \label{fig:e2e}
\end{figure*}

\section{DelightfulTTS 2}

As shown in Figure~\ref{fig:e2e}, our proposed DelightfulTTS 2 consists of two components: 1) a codec network based on vector-quantized auto-encoders with adversarial training (VQ-GAN) that encodes raw waveform into frame-level feature embeddings with its encoder and quantizer and reconstructs waveform with encoded features with its decoder, and 2) an acoustic model based on DelightfulTTS~\cite{liu2021delightfultts} that predicts encoded features from phoneme sequence. A joint training strategy for acoustic model and codec is proposed for better voice quality.

\subsection{Speech Representation Learning with VQ-GAN}
\label{sec_codec}

\begin{figure*}[h]
  \centering
  \includegraphics[width=0.93\textwidth,trim=0.5cm 0.5cm 0.5cm 0.5cm,clip=true]{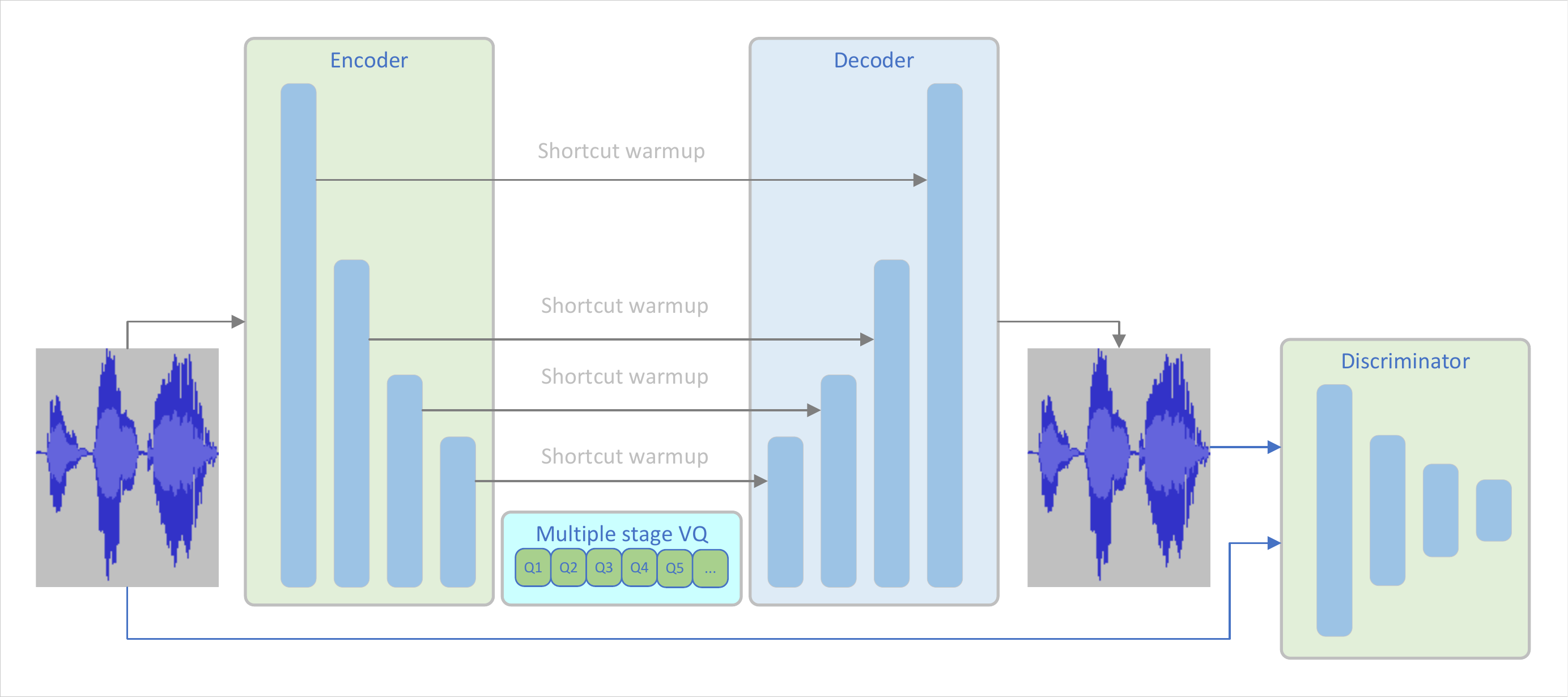}
  \caption{The structure of codec network based on VQ-GAN with multi-stage quantizer. The encoder and multi-stage quantizer convert audio into quantized frame-level speech representations, and the decoder reconstructs audio from the quantized frame-level speech representations with adversarial training.}
  \label{fig:codec}
\end{figure*}

To learn better speech representations instead of mel-spectrograms, we design a new codec network which learns frame-level speech representations based on vector-quantized auto-encoders with adversarial training (VQ-GAN). Specifically, it consists of a symmetric encoder-decoder network ~\cite{shen2018natural,mustafa2021streamwise,zeghidour2021soundstream,tagliasacchi2020seanet} with skip-connections between bottom and top layers, and a multi-stage vector quantizer~\cite{juang1982multiple} in the middle as feature bottleneck. The overall structure of VQ-GAN based codec network is illustrated in Figure~\ref{fig:codec}.

The decoder adopts the same architecture as the generator in HiFi-GAN ~\cite{kong2020hifi}, with a bidirectional Long Expressive Memory (LEM) ~\cite{rusch2021long} layer at upsampling stage for a stable learning of long-term sequential dependencies. The encoder mirrors the decoder in its layout. Skip-connections are added between the first three encoder blocks and its mirrored decoder blocks during training, which we found is critical to stabilize joint training and help model convergence. Multi-stage vector quantization~\cite{juang1982multiple} is applied on top of the codec encoder, to quantize each frame of encoded features in multiple stages.

For adversarial training, we combine the same multi-scale and multi-period discriminators as ~\cite{kumar2019melgan,kong2020hifi}. Three discriminators with the same structure are applied to input audio at different resolutions: original, 2x down-sampled, and 4x down-sampled. Discrete wavelet transform is used to replace average sampling in the discriminators as ~\cite{kim2021fre} to reproduce high-frequency components accurately.

\subsection{Acoustic Model based on DelightfulTTS}
\label{sec_am}
The acoustic model predicts the quantized speech representations with phoneme sequence as input. The network is based on DelightfulTTS~\cite{liu2021delightfultts}, which consists of an encoder and a decoder with improved conformer blocks, and a variance adaptor to provide multiple variance information to ease the one-to-many mapping between text and speech. The encoder converts a phoneme sequence into its hidden representations, and the variance adaptor predict the variance information including utterance-level acoustic condition, phone-level acoustic conditions, and phoneme-level pitch and duration, and then the decoder predicts the frame-level speech representations with variance information and phoneme hidden as input.

\subsection{Joint Training of Acoustic Model and Vocoder}
\label{sec_joint}
Previous two-stage cascaded TTS systems consist of an acoustic model and a vocoder that are trained independently.
Although these models can synthesize speech with good quality, they have a few drawbacks:
Firstly, there is a feature mismatch between training and inference phase for the input of vocoder, i.e., ground-truth mel-spectrograms in training while predicted ones in inference.
Secondly, pre-designed features like mel-spectrograms limit the performance of waveform reconstruction, which loses phase information and high-frequency details.
In DelightfulTTS 2, we jointly train the acoustic model and vocoder in an end-to-end way to improve TTS performance, with a specifically designed scheduled sampling mechanism in acoustic model.
\begin{itemize} [leftmargin=*]
\item The acoustic model~\cite{liu2021delightfultts} has four variance information modules: duration predictor, pitch predictor, utterance-level acoustic predictor and phone-level acoustic predictor. During training, the ground-truth pitch, utterance-level acoustic embedding, and phone-level acoustic embedding are extracted from ground-truth mel-spectrograms and added to phoneme hidden as decoder input to predict speech representations.
This poses a mismatch between training and inference: predicted variance information in inference stage has gap compared to ground-truth counterparts. Instead of feeding all ground-truth features during training, we leverage a schedule sampling mechanism~\cite{liu2020teacher} for pitch, utterance-level acoustic condition, and phone-level acoustic conditions, to reduce training and inference gap, and improve the performance of end-to-end model in test phase.
\item During training phase, acoustic model output is directly used as vocoder input, with an auxiliary L1 loss between acoustic model output and quantized speech representations to stabilize training, and then a random segmentation process is applied on top of the predicted representations.
\end{itemize}

\subsection{Training Objectives}
\textbf{Discriminator Loss} The adversarial objectives for VQ-GAN and end-to-end training follow ~\cite{kong2020hifi}, with multi-period discriminator loss and multi-scale discriminator loss. To alleviate high-frequency loss, we further replace problematic average pooling method with discrete wavelet transform ~\cite{nason1994discrete, kim2021fre} as the down sampling method, which is an effective way of down sampling non-stationary signals into several frequency sub-bands.
\\\textbf{Codec Decoder Loss} Codec decoder (vocoder) in end-to-end training has a multi-resolution spectrogram loss $L_{mrs}$, adversarial loss $L_{\rm Adv}$ and feature match loss $L_{\rm fm}$ ~\cite{yamamoto2020parallel, kumar2019melgan}, which help to generate realistic results when jointly optimizing with adversarial loss functions. $L_{\rm vq}$ is vector quantization loss ~\cite{srikotr2021vector} for all vector quantizers.
\begin{equation}
L_{\rm G} = L_{\rm Adv}+L_{\rm vq}+L_{\rm fm}+L_{\rm mrs}
\end{equation}
\\\textbf{Acoustic Model Loss} Following ~\cite{liu2021delightfultts}, we introduce phoneme-level pitch and duration loss, utterance-level, and phoneme-level acoustic condition loss, where $L_{\rm utt}$/$L_{\rm phone}$ is the L1 loss between predicted utterance-level/phoneme-level acoustic condition vector and the vector extracted from utterance-level/phoneme-level reference encoder; $L_{\rm pitch}$/$L_{\rm dur}$ is the L1 loss between predicted pitch/duration and the ground-truth pitch/duration. To improve audio fidelity, we use SSIM ~\cite{wang2004image} to measure the similarity between predicted by the acoustic model and ground-truth quantized speech representations by the codec encoder, denoted as $L_{\rm ssim}$. $L_{\rm feat}$ is the L1 loss between predicted speech representations and quantized speech representations.
\begin{equation}
L_{\rm AM} = L_{\rm pitch}+L_{\rm dur}+L_{\rm utt}+L_{\rm phone}+L_{\rm SSIM}+L_{\rm feat}
\end{equation}
The overall joint training loss function for DelightfulTTS 2 is a combination of acoustic model and audio codec decoder loss, $W_{\rm G}$ and $W_{\rm AM}$ are the loss weights of different components.
\begin{equation}
L_{\rm joint} = W_{\rm G}*L_{\rm G}+W_{\rm AM}*L_{\rm AM}
\end{equation}

\section{Experiments and Results}
In this section, we conduct experiments to evaluate the effectiveness of the proposed DelightfulTTS 2. We first describe the experimental setup and then introduce the results of DelightfulTTS 2. Audio samples are available at \footnote{Audio samples: \url{https://cognitivespeech.github.io/delightfultts2}}.

\subsection{Experimental Setup}
\textbf{Datasets} We conduct experiments on an internal dataset with 40-hour professional English speech-script pairs. This dataset is divided into train set, dev set, and test set respectively. All text in our datasets are first processed by text normalization (TN) with a rule-based TN module and then converted into phonemes by a grapheme-to-phoneme module. The duration target is extracted by an internal force alignment model. The frame-level speech representation is obtained by downsampling audio samples by the codec encoder, which has a 12.5 ms hop size like mel-spectrogram ~\cite{liu2021delightfultts}. The mel spectrogram used in our system is extracted from audio downsampled to 16 kHz and computed through a short time Fourier transform (STFT) using a 50 ms frame size, 12.5 ms frame hop. Frame-level pitch is extracted on 16KHz speech too and then averaged to phone-level pitch with phoneme alignment ~\cite{liu2021delightfultts}. For codec training, the original 48 kHz audios were downsampled to 24 kHz.
\\\textbf{Training Setting} Our training process involves first training the codec network on audio samples only, followed by jointly training an acoustic model with the pretrained codec decoder (vocoder) model. To train the codec network, we apply the standard exponential moving average with a batch size of 16 on eight 32GB V100 GPUs, with a random audio segment of 24000 waveform points. The shortcut between codec encoder and decoder will be removed after the first 1k warm up steps to accelerate convergence. Adam optimizer ~\cite{kingma2014adam} is used, and a learning rate of 0.0001 with exponentially decaying to 0.0001 starting from 100,00 iterations. 
\\\textbf{Evaluation Setting} We keep an evaluation set randomly preserved from the training set and evaluate the audio quality with both subjective and objective metrics. The evaluation set covers long sentences, short sentences, question sentences, and exclamation sentences, etc. Subjective metrics include mean opinion score (MOS) and comparative mean option score (CMOS) using the same judgement system and group. Audio generated on this set is sent to a human rating system where each sample is rated by at least 20 raters on a scale from 1 to 5 with 0.5-point increments, to better verify TTS system performance, we filtered the judges with same or better TTS score over recording. For each pair of utterances with a random order in CMOS, 20 raters are asked to give a score ranging from -3 (new system is much worse than baseline) to 3 (new system is much better than baseline) with intervals of 1. For objective metrics, we report results measured by means of ViSQOL ~\cite{hines2015visqol} in ablation studies.

\subsection{Results}
\label{sec_test_codec}
\textbf{Speech Quality} Table~\ref{tab:mos} shows a comparison of our method against baselines. DelightfulTTS and FastSpeech 2 use phoneme sequence as input and mel-spectrogram as target, and ~\cite{ms2020hifinet} synthesizes waveform with mel-spectrogram. Results show that the proposed system outperforms the baseline systems in terms of naturalness.

\begin{table}[h]
  \centering
  %\resizebox{\linewidth}{!}{%
  \begin{tabular}{lll}
    \toprule
    System & MOS \\
\midrule
    Ground Truth & 4.39 ± 0.08 & \\
  \midrule
  FastSpeech  2~\cite{ren2021fastspeech} & 4.08 ± 0.09 & \\
 
    DelightfulTTS~\cite{liu2021delightfultts}  & 4.16 ± 0.09 & \\
    \midrule
    DelightfulTTS 2 & 4.26 ± 0.09 & \\

    \bottomrule
  \end{tabular}
  \caption{ MOS evaluations of different systems.}
  \label{tab:mos}
\end{table}

\noindent We also conduct a side-by-side evaluation in Table~\ref{tab:cmos} between audio synthesized by DelightfulTTS 2 and other systems, from which a comparison mean option score (CMOS) is calculated. The overall mean scores of +0.14 over DelightfulTTS and +0.13 over FastSpeech 2 show that raters have a statistically significant preference towards our system over other TTS systems; mean score of -0.06 over recording indicates that voice quality of DelightfulTTS 2 has slight regression compared to recording in general domain speaking of naturalness.  

\begin{table}[h]
  \centering
  %\resizebox{\linewidth}{!}{%
  \begin{tabular}{lll}
    \toprule
    Baseline & CMOS \\
    \midrule
    FastSpeech 2 & +0.13 & \\
    \midrule
    DelightfulTTS & +0.14 & \\ 
    \midrule
    Recording & -0.06 & \\ 
    \bottomrule
  \end{tabular}
  \caption{CMOS Comparison of DelightfulTTS 2 (new system) vs different TTS systems and recording. }
  \label{tab:cmos}
\end{table}

\noindent \\\textbf{Inference Latency} Acoustic model has about 65.3M parameters and vocoder has about 3.3M parameters. We evaluate the inference latency of DelightfulTTS 2 on NVIDIA V100 GPU, the end-to-end RTF is about 0.008, which is on par as DelightfulTTS.

\subsection{Analyses on Codec Network }
\label{sec_codec_result}
To test the reconstruction performance of the proposed audio codec, both subjective and objective metrics are involved. The test set is the evaluation set persevered from training set. For subjective test, a side-by-side CMOS evaluation with 20 raters between audio synthesized by audio codec (new system) and ground truth (baseline systems) is shown in Table~\ref{tab:re_cmos}. The overall mean score of -0.03 shows that the waveform reconstructed by our proposed codec network has no obvious significance with the ground-truth waveform.

\begin{table}[h]
  \centering
  %\resizebox{\linewidth}{!}{%
  \begin{tabular}{lll}
    \toprule
    System & CMOS \\
    \midrule
    Codec Reconstruction vs Ground Truth & -0.03 & \\
    \bottomrule
  \end{tabular}
  \caption{CMOS for codec reconstruction.}
  \label{tab:re_cmos}
\end{table}

\noindent For objective quality metrics, Table~\ref{tab:obj1} shows the rate-quality curve of proposed codec network over a wide range of bitrates, from 3.2 kbps to 12.8 kbps. As measured by means of ViSQOL ~\cite{hines2015visqol}, we observe that the reconstructed speech quality by codec decoder decreases as the bitrate is reduced; on the other hand, the number of the intermediate speech frame also decreases as the bitrate is reduced, which may be easier for the acoustic model to predict and faster for runtime inference. Our proposed codec operates at constant bitrate 12.8 kbps, resulting in the same frame length as traditional mel-spectrogram, to balance efficiency of end-to-end TTS modelling and reconstructed speech quality. Given bitrate 1.1 kbps, Table~\ref{tab:obj2} shows the ViSQOL for different frame-level speech representation dimension extracted by codec encoder, which indicates that larger dimension has slightly better quality.

\begin{table}[h]
  \centering
  %\resizebox{\linewidth}{!}{%
  \begin{tabular}{llll}
    \toprule
    Bitrate & 3.2 & 6.4 & 12.8 \\
    \midrule
    ViSQOL & 3.80 & 3.90 & 4.28\\
    \bottomrule
  \end{tabular}
  \caption{Comparison of ViSQOL for different bitrate (kbps).}
  \label{tab:obj1}
\end{table}

\begin{table}[h]
  \centering
  %\resizebox{\linewidth}{!}{%
  \begin{tabular}{lllll}
    \toprule
    Dim & 256 & 512 & 1024 & 2048 \\
    \midrule
    ViSQOL & 3.39 & 3.41 & 3.51 & 3.55\\
    \bottomrule
  \end{tabular}
  \caption{Comparison of ViSQOL for different frame-level speech representation dimension under 1.1 kbps.}
  \label{tab:obj2}
\end{table}

\section{Conclusions}
This paper describes DelightfulTTS 2, an end-to-end TTS system that combines a convolutional codec network with adversarial vector-quantized auto-encoders and an acoustic model based on DelighfulTTS. DelightfulTTS 2 can be trained jointly from paired text/audio data, without suffering from cascaded errors in two-stage TTS models and sub-optimal pre-designed acoustic features like mel-spectrograms. The synthesized speech of DelightfulTTS 2 achieves better quality than baseline systems and is of similar quality to recordings. The model parameters count and synthesis speed are similar to DelightfulTTS. For future work, we plan to investigate more efficient codec approaches to improve TTS model performance on more challenging datasets such as multi-speaker or multi-lingual datasets, and the potential solutions for low latency speech syntheses application like device or streaming.

\bibliographystyle{IEEEtran}

\bibliography{mybib}

\end{document}